\documentstyle[prb,aps,twocolumn,epsf]{revtex}

\begin{document}
\draft

\title{Ballistic electron transport exceeding 160~$\mu$m in an undoped
	GaAs/AlGaAs FET}
\author{G. R. Facer\cite{email}, B. E. Kane, A. S. Dzurak, R. J. Heron,
	N. E. Lumpkin and R. G. Clark}
\address{Semiconductor Nanofabrication Facility,\\  University of
	New South Wales, Sydney NSW 2052  AUSTRALIA}
\author{L. N. Pfeiffer and K. W. West}
\address{Bell Laboratories, Lucent Technologies,
         Murray Hill NJ 07974 USA \\}
\date{\today}
\maketitle

\begin{abstract}

We report measurements of GaAs/AlGaAs undoped field-effect
transistors in which two-dimensional electron gases (2DEGs) of exceptional
quality and versatility are induced without modulation doping.  Electron
mobilities at $T=4.2$~K and density $3 \times 10^{11}$~cm$^{-2}$ exceed
$4 \times 10^{6}$~cm$^{2}$V$^{-1}$s$^{-1}$.  At lower temperatures,
there is an unusually large drop in scattering, such that the mobility
becomes too high to measure in 100~$\mu$m samples.  Below $T=2.5$~K, clear
signatures of ballistic travel over path lengths in excess of 160$~\mu$m are
observed in magnetic focusing experiments.  Multiple reflections at the edges
of the 2DEG indicate a high degree of specularity.\\

\end{abstract}



High quality two-dimensional electron systems in GaAs/Al$_x$Ga$_{1-x}$As
heterostructures form the basis for most semiconductor mesoscopic devices,
because they have for some time been used successfully
to achieve long mean free paths in the electron layer.  To date,
modulation-doped (MD) structures, where dopants spatially separated from
the GaAs/AlGaAs junction induce a two-dimensional electron gas (2DEG), have
yielded the highest electron mobilities\cite{UmanskyHM,LNP_conv}.

In 2DEGs where the mean free path becomes comparable to the device
dimensions, oscillations in four-terminal resistance measurements can be
seen in small magnetic fields $B$.  The $B$-periodic component corresponds
to the electron cyclotron orbit diameter passing through integer divisors
of the sample size.  This classical phenomenon, referred to as electron
focusing\cite{vanHouten_foc,Foxon_IBM}, relies on scattering-free
or {\em ballistic} transport across the device.  Extensive work has been
performed on electron focusing in 2DEGs over a wide range of device
sizes ($<1$~$\mu$m to 150~$\mu$m) \cite{Spector_SS1990,HirayamaAPL_1991}.
In addition to the many experiments which have been performed to investigate
striking behaviour such as fractal self-similarity in the magnetoresistance
of very small samples\cite{Taylor_billiard}, and the development of
ballistic electron optics\cite{Spector_SS1992,Spector_APL}, electron mobility
improvement through further development of epitaxial growth techniques has
resulted in the observation of focusing effects up to length scales of order
150~$\mu$m\cite{HirayamaAPL_1991}.

Saku {\em et al} have predicted\cite{Saku_lim} that the highest obtainable
mobility in MD systems with electron densities
$N \sim 3 \times 10^{11}$~cm$^{-2}$ is likely to be
$\approx 1.6 \times 10^7$~cm$^2$V$^{-1}$s$^{-1}$.  This prediction is based
on scattering from the potential landscape contributed by randomly located
ionized dopants themselves.  However, recent experimental studies on MD
structures\cite{UmanskyHM,LNP_conv} have presented strong evidence that the
effect of dopant disorder is not as severe as expected from that model.
Whether or not the inherent dopant disorder is the limiting factor for the
mobility, there is a way to avoid its effects:  the 2DEG can be
induced by a gate bias in an accumulation-mode undoped field-effect transistor
(UFET)\cite{BK_FET,Herfort_UFET}. As the name implies, UFETs contain a high
quality, variable-density 2DEG without the use of modulation doping.  We
report magnetoresistance measurements in several UFETs, where a series of
sharp magnetic focusing peaks demonstrate ballistic
transport across 100~$\mu$m devices over the wide density range
$N= 2 \times 10^{11}$~cm$^{-2}$ to $6 \times 10^{11}$~cm$^{-2}$, with
ballistic scattering lengths exceeding 160~$\mu$m established in small
magnetic fields at high densities.

\begin{figure}
\epsfxsize=6cm
\vspace{-8cm}
\centerline{\epsfbox{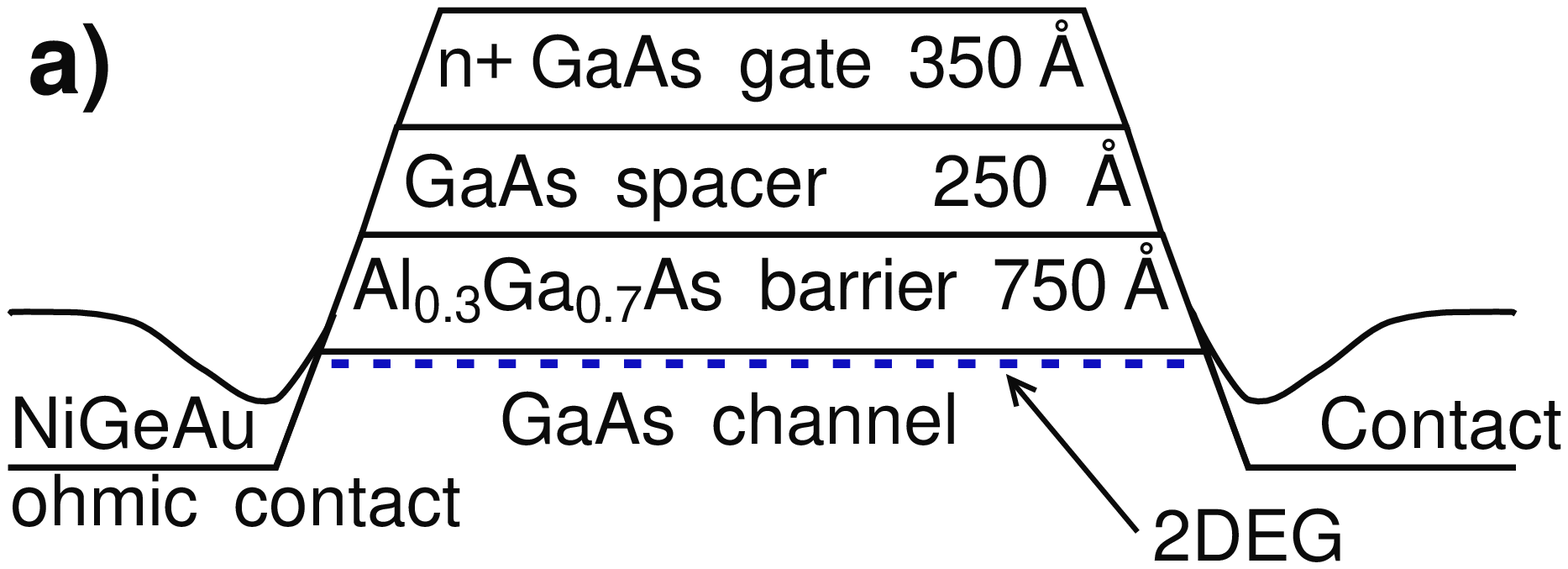}}
\vspace{-3cm}
\epsfxsize=6.5cm
\centerline{\epsfbox{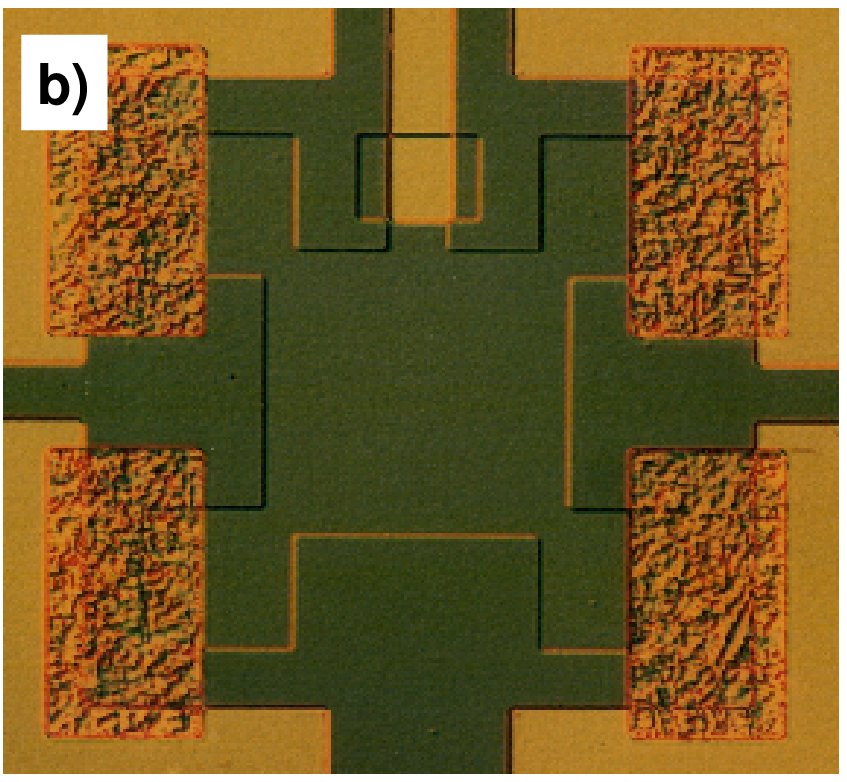}}
\epsfxsize=6cm
\vspace{-0.6cm}
\centerline{\epsfbox{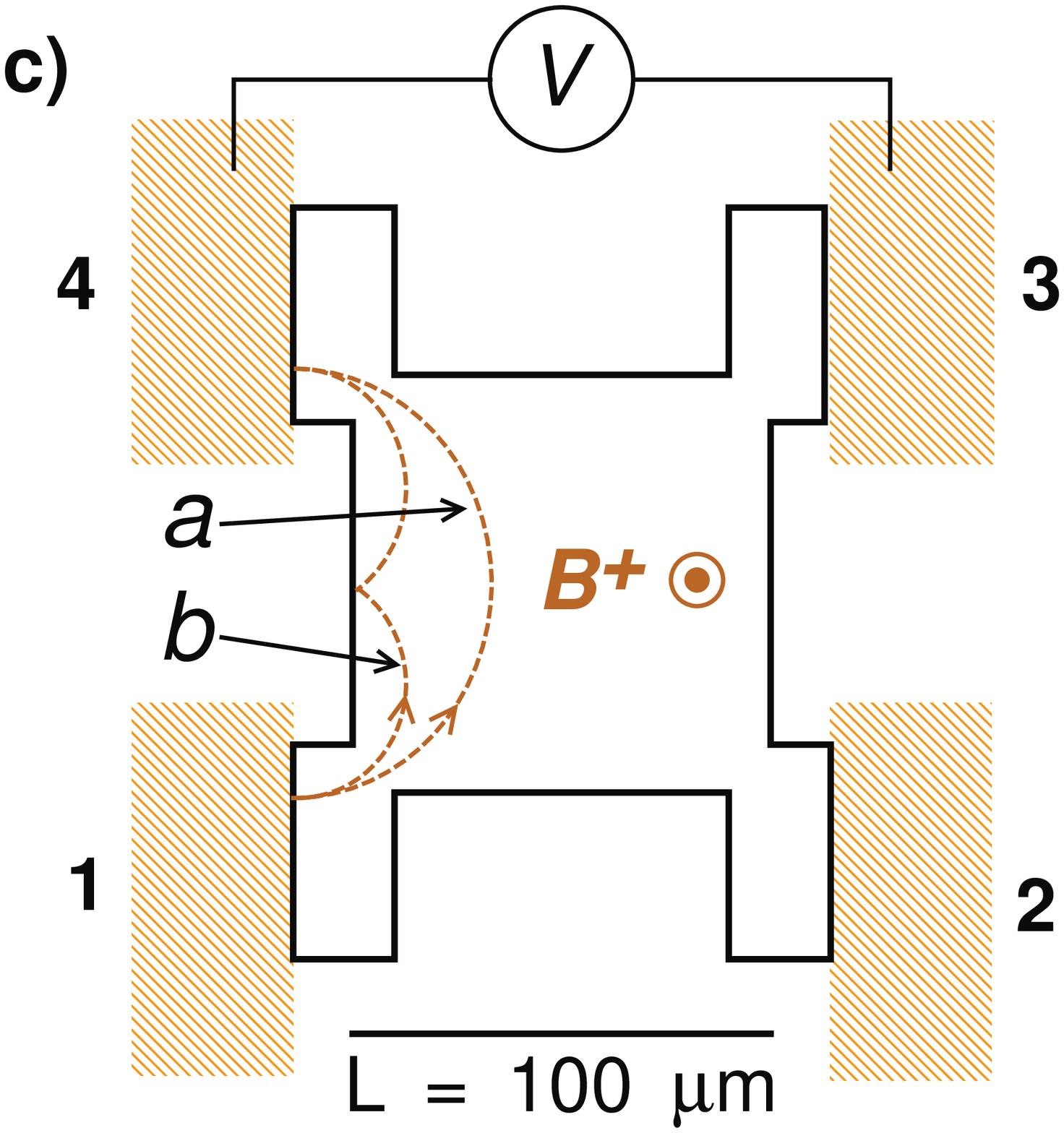}}
\caption{
(a)  Layer structure of the FET devices.  (b)  Optical microscope picture of
a UFET device:  the width of the central square is 100~$\mu$m.  Mottled
regions in the corners are self-aligned ohmic contacts which make contact to
the electron layer without a short circuit to the top gate.  (c)  Schematic
diagram, defining the labelling convention for the current
$I_{1 \rightarrow 2}$, and the measured voltage $V_4 - V_3$, between
ohmic contacts.
}
\label{devices}
\end{figure}

The 2DEGs under investigation form at a GaAs/AlGaAs
heterojunction upon applying a positive bias to an epitaxially-grown top
gate\cite{BK_FET} (Fig.~\ref{devices}(a)).  The
lack of dopant-induced disorder greatly reduces scattering, in particular
the small-angle scattering from distant ions\cite{DasSarma_tauPRB}.
  UFETs also allow precise tuning of the electron sheet density over
nearly two orders of magnitude.

Optical lithography was used to fabricate UFETs configured as squares
with side length 100~$\mu$m (Fig.~\ref{devices}(b,~c)).  The 2DEG is
measured via self-aligned ohmic contacts\cite{BK_FET}.  Van der Pauw
resistance measurements have been performed in a dilution refrigerator using
low-frequency ($\sim 15$~Hz) lock-in techniques with currents less than
150~nA.  The temperature-dependent data were taken with gate leakage less
than 300~pA.  Practical device sizes in the wafers reported here are limited
to $\sim$~100~$\mu$m by the presence of oval defects which can cause a
short circuit between the gate and the 2DEG.

\begin{figure}
\epsfxsize=8cm
\vspace{-2cm}
\centerline{\epsfbox{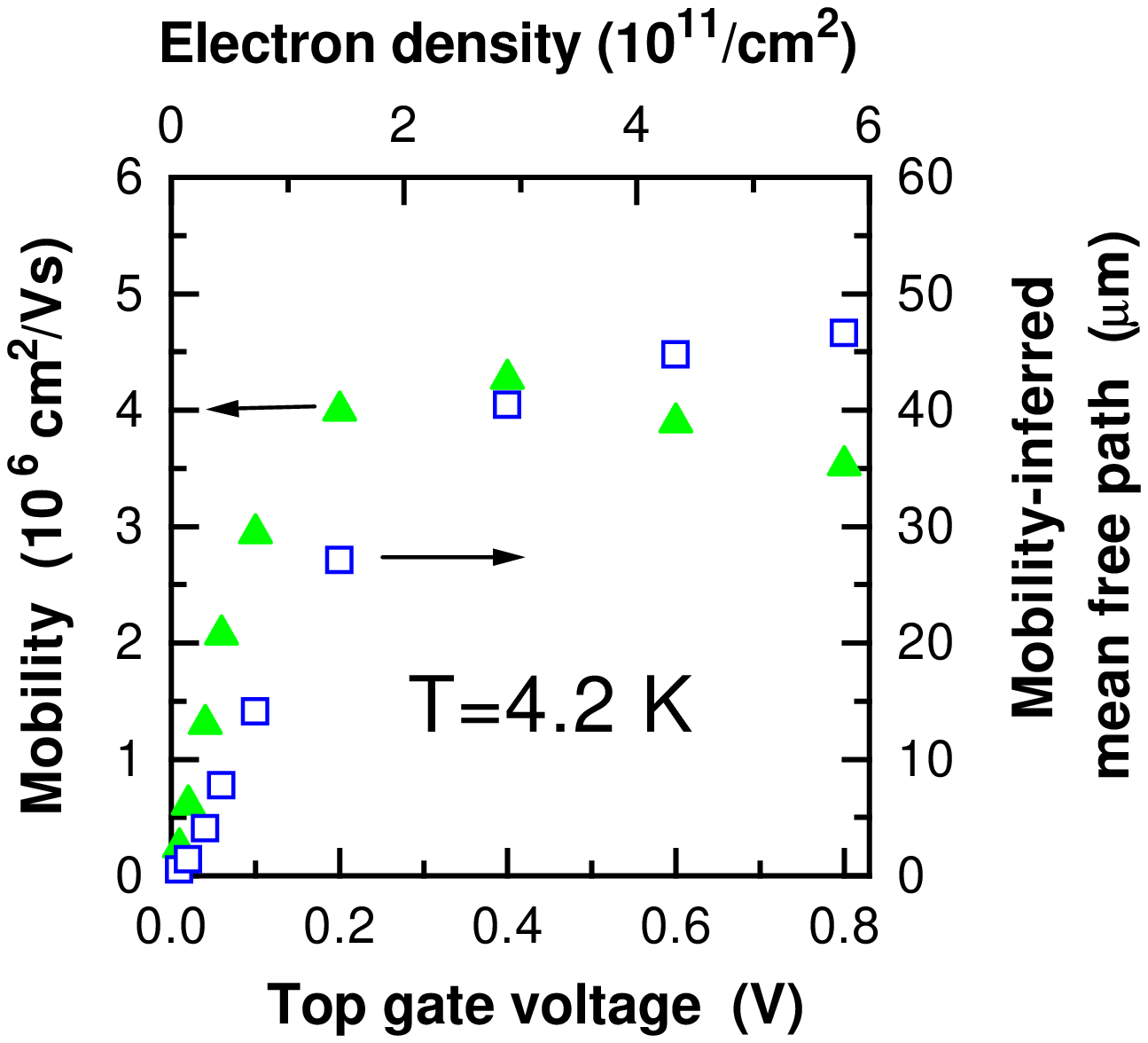}}
\caption{Electron sheet density $N$, mobility $\mu$ (filled triangles),
and mean free path (squares) calculated from $N$ and $\mu$, at $T=4.2$~K,
for a UFET.
}
\label{4Kprop}
\end{figure}

Electron density and mobility at $T = 4.2$~K are shown in
Fig.~\ref{4Kprop} as a function of top gate voltage.  The electron
mobility peaks at a density $N \approx 3 \times 10^{11}$~cm$^{-2}$,
an effect which is more pronounced at temperatures below $T = 1$~K.
The presence of this mobility peak is believed to be due to the increasing
influence of interface roughness at higher gate voltages, as the 2DEG is
pulled closer to the GaAs/AlGaAs heterojunction.  The scattering lengths
given for the 4.2~K data in Fig.~\ref{4Kprop} are simply determined
from the mobility.  The highest values are of order half the sample size.

Below $T = 4.2$~K, the electron mobility is expected to improve due to
the reduction of electron--phonon scattering.  If the mean free path
exceeds the device dimensions, then a meaningful sheet resistivity cannot
be defined, and thus a conventional electron mobility cannot be determined.
Instead, the scattering length of the electrons must be used to describe
the quality of the 2DEG system.

\begin{figure}
\epsfxsize=10cm
\hspace{2.5cm}
\centerline{\epsfbox{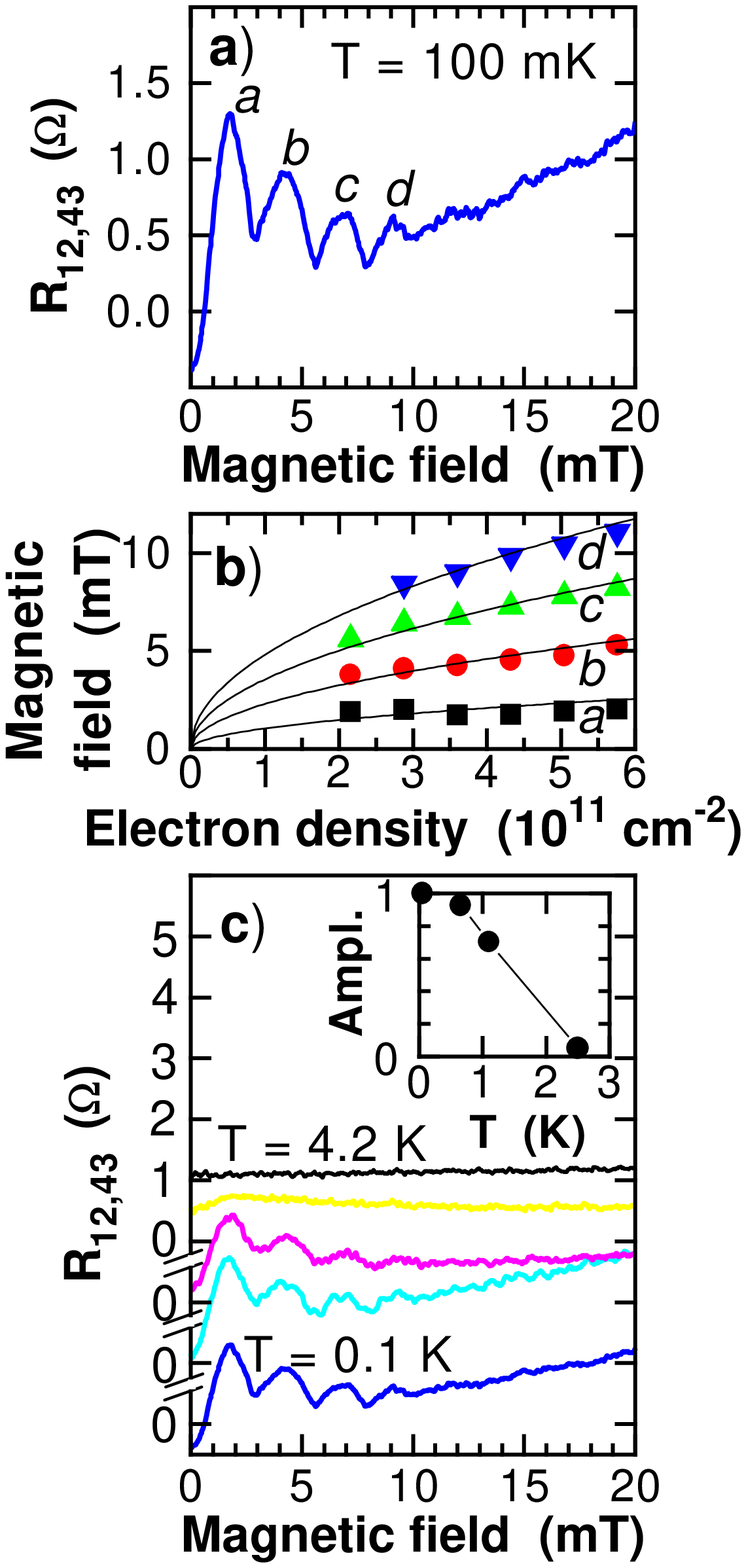}}
\caption{(a) Magnetoresistance for $B \geq 0$.  (b) Oscillation positions as
a function of electron density.  The solid curves are proportional to the
square root of the electron density.  (c) Oscillations at a range of
temperatures: $T =$~0.1, 0.65, 1.1, 2.5, and 4.2~K.  Inset: graph of
oscillation amplitude, defined as the change in resistance from the central
minimum to the first $B^+$ peak, vs $T$.
}
\label{oscs}
\end{figure}

In the UFET samples, for $T<4.2$~K, magnetoresistance oscillations are visible
at fields on the order of 0.01~T, shown in Fig.~\ref{oscs}(a).
The vertical axis is the four-terminal measurement\cite{Buttiker}
$R_{12,43} = (V_4 - V_3)/I_{1 \rightarrow 2}$.  Near $B=0$, the measurement
can actually be negative, as seen in Fig.~\ref{oscs}(a), because electrons
are injected directly along the diagonal of the device into the negative
voltage-measuring contact $V_3$\cite{HirayamaAPL_1991,Buttiker}.
Focusing--related oscillations are expected to be periodic in $B$ at any
given density, and the field position $B_{foc}$ of the first peak scales as
the square root of $N$:
\begin{equation}
B_{foc} = \frac{2 \hbar \sqrt{2 \pi N}}{eL},
\label{periodeqn}
\end{equation}
where $L$ is the size of the device.  The experimentally observed
oscillations {\em are} $B$-periodic, and the wide tunability of
the UFET electron density allows analysis of the density
dependence.  Fig.~\ref{oscs}(b) gives the oscillation positions (maxima)
as a function of electron density.  The solid curves in the figure show
that the periods are consistent with an increase as the square root of $N$.
Note the large magnitude of the oscillations relative to the background
signal;  no subtraction of a background level is necessary for this data.
  The number of visible oscillations is related to the specularity of
reflections at the edges of the UFET, rather than to the scattering
lengths within the 2DEG region.  Specularity becomes relevant because
each successive peak corresponds to an added skipping orbit.  Observation of
several consecutive oscillations (labeled {\em a, b, c, d} in Fig.~\ref{oscs})
indicates that the reflections from the UFET boundaries are highly specular.

Oscillations at a range of temperatures, normalized to the
base temperature limit, are shown in Fig.~\ref{oscs}(c).  Notably, there is a
significant temperature dependence below $T=1$~K, which is stronger than
the mobility change, dominated by electron--phonon scattering, usually
observed in MD structures\cite{Stormer_phonon,Mendez_phonon,LNP_growth}.

\begin{figure}
\epsfxsize=8cm
\centerline{\epsfbox{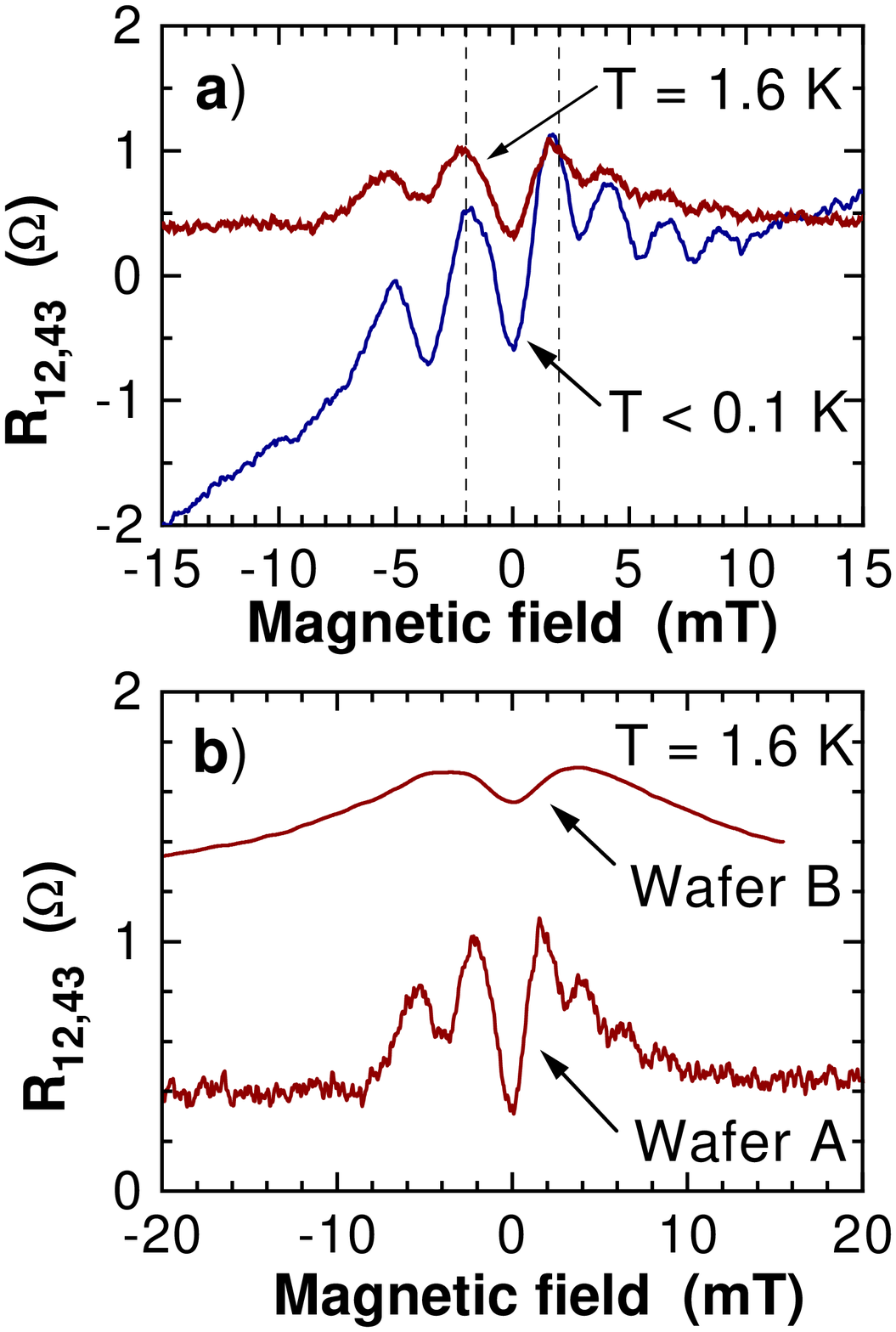}}
\caption{(a) Magnetoresistance spanning $B = 0$, in the configuration of
Fig.~\ref{devices}(c).  Vertical, dotted lines indicate the expected positions
of the first $B^+$ and $B^-$ focusing peaks for a 100~$\mu$m square.  (b)
Comparison between two samples (from different wafers) at the same electron
density.
}
\label{ltgt0}
\end{figure}

Magnetoresistance data spanning both field polarities ($B^+$ and $B^-$) are
shown in Fig.~\ref{ltgt0}(a).  For a diameter
of 100~$\mu$m, Eq.~(\ref{periodeqn}) predicts a position of the first peak
$B_{foc}=1.98$~mT. The vertical, dotted lines in Fig.~\ref{ltgt0}(a)
mark $B = \pm$1.98~mT, and it can be seen that the
experimental data agree very well with the predicted $B_{foc}$.
The electron focusing peaks therefore imply that the ballistic mean free
path is at least $(\pi /2 \times~100) \approx 160$~$\mu$m.  Indeed, given
the strength of the oscillations (recalling that there is no background
subtraction), the path length probably exceeds this value,
but explicit confirmation requires larger samples which are constrained by
the oval defects in these high-mobility wafers, described earlier.

  At the lowest temperatures, there is an antisymmetric background slope
near $B=0$, which is visible only for $|B| \lesssim 100$~mT.  This slope is
believed to be related to the Hall effect, and is evident even in the
standard four-terminal geometry because of the ballistic nature of transport
in the UFETs.

Although the observed position of the first $B^+$ and $B^-$ focusing peaks
agrees well with the predicted $B_{foc}$, the subsequent period $\Delta B$
separating successive oscillations is larger than $B_{foc}$ by 20~\%.  A
larger $\Delta B$ indicates a tighter (i.e. {\em smaller}-diameter) cyclotron
orbit than is expected for the 100~$\mu$m devices.  While there are several
possible reasons why $\Delta B$ could be greater than $B_{foc}$,
such as the lack of four-fold symmetry in the device or anisotropy of
injection/absorption of electrons at the self-aligned ohmic contacts (which
have resistances $\sim 2$~k$\Omega$), a convincing explanation of why
$\Delta B > B_{foc}$ is yet to be found.

Fig.~\ref{ltgt0}(b) shows data from UFETs made from two wafers.  The trace
at higher resistance (labeled ``Wafer B'') is data taken as part of a
far-infrared photoconductivity study\cite{RJH_FIR} which has yielded a
very narrow cyclotron resonance linewidth (full width at half maximum) of
6~mT at $T=1.6$~K and at a wavelength of 183.4~$\mu$m.  The wafer B sample
has the lower $T=4.2$~K mobility of $1 \times 10^6$~cm$^{2}$V$^{-1}$s$^{-1}$,
for $N=3.6 \times 10^{11}$~cm$^{-2}$.  The ballistic transport oscillations
in the ``Wafer A'' trace are much better resolved.  The different form
of the traces from 2DEGs of unequal quality also highlights
another aspect of the ballistic transport data: the $B=0$ resistance minimum,
used in itself in previous studies\cite{HirayamaAPL_1991} as
evidence for ballistic transport for which sequential peaks are barely
resolved, is far more robust than the later focusing peaks, which degrade
to a broader envelope.

Several length scales have been used to describe scattering in high-mobility
semiconductor heterostructures.  To minimize confusion, we employ the
convention of Spector {\em et al.}\cite{Spector_SS1990,Spector_SS1992}.  The
most commonly used distance is the elastic mean free
path $\lambda_{\mu}$, which is the distance derived from mobility calculations.
$\lambda_{\mu}$ is the distance over which an electron travels before its
momentum is completely randomized.  Next, there is a ballistic mean free
path $\lambda_{b}$ which is the distance over which  ballistic focusing peaks
are visible.  $\lambda_{b}$ is more sensitive to small-angle scattering than
$\lambda_{\mu}$, since several small-angle scattering events are required to
fully randomize the motion of an electron, whereas ballistic focusing
relies fundamentally on directional effects.  In the case of these UFETs,
there will be a small angular tolerance introduced by the finite width of
the corner constrictions (Fig.~\ref{devices}(b)).  Finally, there is
$\lambda_{q}$, the length which is obtained via the quantum lifetime from
Shubnikov-de-Haas oscillations\cite{AFS}.  $\lambda_{q}$ is expected to be
the smallest of all the various path lengths\cite{AFS};  in GaAs/AlGaAs
systems, generally $\lambda_{\mu}>\lambda_b>\lambda_q$.

Related to $\lambda_{q}$ is the lifetime obtained via cyclotron resonance
measurements.  The 6~mT resonance linewidth stated above for wafer B
yields a scattering lifetime of $130$~ps in a low density regime
($N = 2.7 \times 10^{10}$~cm$^{-2}$).  If one na\"{\i}vely multiplies this
lifetime by the Fermi velocity, a path length of 9~$\mu$m is
obtained, attesting to the high quality of even the lower-mobility
wafer B sample.

It is known that the ionized dopants in modulation-doped structures act as
small-angle scatterers\cite{AFS}, in addition to scattering by any random
intrinsic impurities.  In a recent study of modulation-doped layers,
Umansky {\em et al.}\cite{UmanskyHM} report a mobility at $T=0.1$~K of
$1.44 \times 10^7$~cm$^{2}$V$^{-1}$s$^{-1}$.  In that paper, the authors
calculate that the dopants contribute only 10 percent of the total scattering.
  The mobility yields $\lambda_{\mu}$~$\gtrsim$~110~$\mu$m at the stated
$N=2.4\times 10^{11}$~cm$^{-2}$, which we note is smaller than the
$\lambda_b$ of 160~$\mu$m in wafer A above.  A previous study of $\lambda_{b}$
in MD structures with $\mu = 1.1 \times 10^7$~cm$^{2}$V$^{-1}$s$^{-1}$ also
indicated that the dominant scattering at $T<0.5$~K is large-angle, probably
due to impurities at the heterojunction interface\cite{Spector_SS1992}.  In
the same paper, the ratio of the elastic and ballistic mean free paths was
found to be $\lambda_{\mu} / \lambda_b \approx 6$.  Applying the same factor
to the ballistic mean free path data reported here would imply a mobility
greater than $10^8$~cm$^{2}$V$^{-1}$s$^{-1}$ and $\lambda_{\mu} \sim 1$~mm;
therefore it is likely that the ratio $\lambda_{\mu} / \lambda_b$ in the UFET
samples is less than that in MD structures.

A mechanism which may be relevant to the $T$-dependence of the focusing peak
amplitudes is electron--electron interaction.  Electron--electron scattering
is expected to depend on $T$ as $T^2\log(T/T_F)$, where $T_F$ is the Fermi
temperature\cite{Giuliani_ee}.  While it is not expected that
electron--electron interactions should have a significant effect on the
resistivity, such interactions have been demonstrated to have an impact on
resistance measurements of smaller devices in a ballistic transport
regime\cite{Molenkamp}.  The amplitude data in the inset to Fig.~\ref{oscs}(c)
do not fit such a temperature dependence well, although a one-to-one
correspondence between the interaction rate and the peak amplitudes should
not necessarily be expected.  The precise mechanism involved in the large
mean free path improvement below $T=4.2$~K is therefore yet to be identified.

Even if mobilities in MD structures improve substantially, UFET samples will
prove of greatest benefit in the production of one- or zero-dimensional
structures.  This is because, in nanostructures, even the minimal fluctuations
caused by dopants (sufficient only to cause small-angle scattering in a 2DEG)
are enough to significantly perturb the potential landscape, and thus the
device behaviour\cite{NixonDavies}.  In addition, mean free paths improve as
$N$ increases, enhancing the worth of the tunability of UFETs to very high $N$:
in MD structures, high densities are attainable only by bringing the dopants
closer to the 2DEG, with a corresponding increase in disorder.  We have
recently reported\cite{BK_QWAPL} ballistic transport in quantum wires
fabricated from the UFET material described in Fig.~\ref{devices}.

In summary, ballistic path lengths of 160~$\mu$m, limited by practical
UFET device sizes, have been demonstrated in GaAs/AlGaAs UFETs free from
dopant disorder. Such long ballistic paths show that non-modulation-doped
FETs can contain 2DEGs of exceptionally high quality.  The fact that ballistic
mean free paths exceed the dimensions of the device mean that conventional
diffusive descriptions of the electron behaviour, in terms of mobility and
resistivity, are no longer appropriate.  A substantial improvement in
scattering length as a function of temperature below $T=4.2$~K, greater than
has been reported in recent high-mobility modulation-doped 2DEGs, has also
been observed.

This work was supported by the Australian Research Council.



\newpage{}

\newpage{}

\newpage{}

\end{document}